\documentclass[
aps,showpacs,preprintnumbers,amsmath,amssymb]{revtex4}
\usepackage{dcolumn}
\usepackage{bm}
\usepackage{graphicx}
\usepackage{epsfig}
\usepackage{mathrsfs}
\numberwithin{equation}{section}

\numberwithin{equation}{section}

\def\Real{\mbox{Re}}      
\def\i{\mbox{\it i\,}}      
\def\sech{\mbox{\,sech}}  
\def\z{{Y}}               
\def\A{{\mathscr{A}}}     
\def\U{{\mathscr{U}}}     
\def\V{{\mathscr{V}}}     
\def\W{{\mathscr{W}}}     

\newcommand{\df}[2]{\displaystyle\frac{#1}{#2}}
\newcommand{\tf}[2]{\textstyle\frac{#1}{#2}}
\newcommand{\Ssf}[2]{\displaystyle\sum_{#1}^{#2}}

\newcommand{\Intn}{\displaystyle\int}
\newcommand{\Lint}[2]{\displaystyle\int\limits_{#1}^{#2}}

\newcommand{\PDD}[2]{\df{\partial #1}{\partial #2}}

\newcommand{\PDDT}[2]{\df{\partial^{2} #1}{\partial #2^{2}}}

\newcommand{\os}[1]{\overline{#1}}

\newcommand{\eps}{\varepsilon}
\newcommand{\B}[1]{\mbox{\boldmath $ #1 $} }

\newcommand{\be}{\begin{eqnarray}}
\newcommand{\en}{\end{eqnarray}}
\newcommand{\no}{\nonumber}
\newcommand{\Quarter}{\mbox{\tiny $\tf{1}{4}$}}
\newcommand{\Half}{\mbox{\tiny $\tf{1}{2}$}}

\begin{document}

\newcommand{\Eqref}[1]{(\ref{#1})}

\title{Formation of three-dimensional surface waves on deep-water using elliptic solutions of nonlinear 
Schr\"odinger equation}

\author{Shahrdad G. Sajjadi$^{\dag\ddag}$, Stefan C. Mancas$^{\dag}$ and Frederique Drullion$^{\dag}$}
\affiliation{$^\dag$Department of Mathematics, ERAU, Florida, U.S.A.\\
$^\ddag$Trinity College, University of Cambridge, U.K.}


\begin{abstract}

A review of three-dimensional waves on deep-water is presented. Three forms of three dimensionality, namely oblique, forced and  spontaneous type, are identified. An alternative formulation for these three-dimensional waves is given through cubic nonlinear Schr\"odinger equation.
The periodic solutions of the  cubic nonlinear Schr\"odinger equation are found using Weierstrass elliptic $\wp$ functions. It is shown that the classification of solutions depends on the boundary conditions, wavenumber and frequency.  
For certain parameters, Weierstrass $\wp$ functions are reduced to periodic, hyperbolic or Jacobi elliptic functions. It is demonstrated that some of these solutions do not have any physical significance.
An analytical solution of cubic nonlinear Schr\"odinger equation with wind forcing is also obtained which results in how groups of waves are generated on the surface of deep water in the ocean. In this case the dependency on the energy-transfer parameter, from wind to waves, make either the groups of wave to grow initially and eventually dissipate, or simply decay or grow in time.

\end{abstract}

\maketitle

\section{Introduction}

Finite amplitude water waves have been studied since
the pioneering work of G.G. Stokes in 1847 \cite{STK47}.  
Over the years very important 
methods (such as the method of inverse scattering) have been developed 
for obtaining exact solutions to water wave problems, for instance, analytical 
solutions to Korteweg-deVries equation for shallow water waves and
the cubic nonlinear
Schr\"odinger equation which describes wave envelopes for slow
modulation of weakly nonlinear water waves.    The
mathematical richness in the field of water waves appears surprising at first
sight, since the governing differential equation 
is simply that of Laplace's equation but, of course, since the boundary
conditions are nonlinear this gives rise to wealth of
problems.  It is also remarkable that in a classical field such
as that of water waves, new physical phenomena are still being
discovered, both theoretically and experimentally, and many open
questions still remain unanswered.

\begin{figure}[htpb]\label{fig1-3dw}
\begin{center}
\includegraphics[width=12cm]{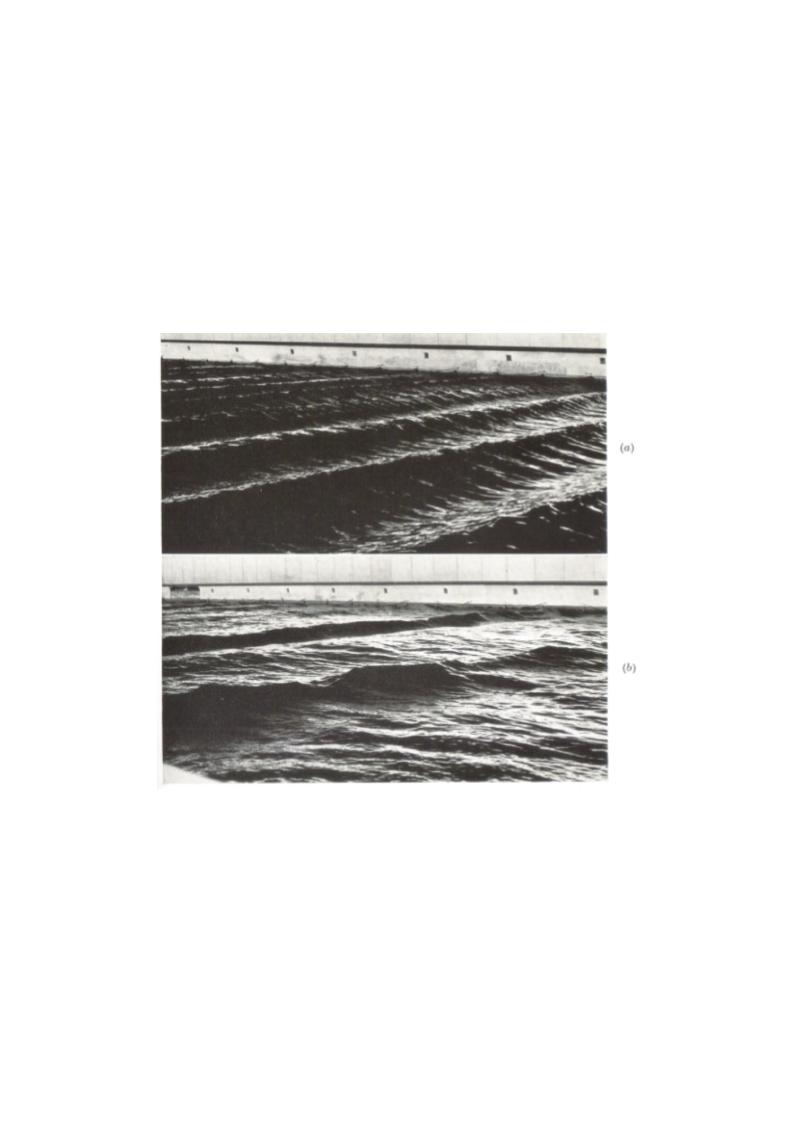}
\caption{Photographs of progressive wave trains, of fundamental wavelength 7.2 ft,
showing disintegration due to Benjamin-Feir instability; (Top) Close to wavemaker; (Bottom) 200 ft
away from wavemaker. From Benjamin \cite{benj67}.}
\end{center}
\end{figure}

For the  two-dimensional periodic, irrotational
surface waves of permanent form propagating under the influence of
gravity on water of infinite or finite depth, such as Stokes waves, much
has been discovered over the last few decades.   For example,
instability of steady finite amplitude waves to long wave
two-dimensional disturbances was first postulated by Lighthill \cite{lighthill} by the
use of Whitham's variational principle, and an approximate Lagrangian.
Zakharov \cite{zakharov} used Hamiltonian methods  and showed that weakly
nonlinear gravity waves are unstable for modulations longer than a
critical wavelength depending upon their waveheight  both analytically and
numerically for two as well as three-dimensional disturbances. Benjamin \& Feir
\cite{bf67a} examined the case of two-dimensional disturbances to weakly
nonlinear waves, by adopting standard perturbation methods, found that 
two-dimensional disturbances, of sufficiently long
wavelength, are unstable for which experimental evidence was demonstrated by 
Benjamin \& Feir \cite{bf67b} and Feir \cite{feir}. In a series of experiments reported by Benjamin \& Feir \cite{bf67b} deep-water wave trains of relatively large amplitude were generated
at one end of a tank, and were seen traveling many wavelengths. They observed that 
these wave trains developed conspicuous irregularities when traveled far
enough, and they completely disintegrated. The severity of the  instability can be
seen in Fig.~1 where the fundamental wavelength is 7.2 ft at the water depth of
25 ft.  Fig.~1~(a) shows the wave train close to the wavemaker, and Fig.~1~(b) 
shows the same wave train at a distance of 200 ft (28 wavelengths) further along the tank.

Extension to Benjamin-Feir instability for three-dimensional oblique waves on 
deep water was first studied by Ross \& Sajjadi \cite{rossaj} using the same perturbation
method as that of Benjamin \& Feir \cite{bf67a}. They discovered a new instability criterion for these waves which reduces to the classical Benjamin-Feir instability
when the angle between the waves is zero \cite{sajrev}. They also demonstrated that in the limiting case,
where the waves are propagating at an angle $\tf{1}{2}\pi$, the instability represents standing waves such as those studied by Penny \& Price \cite{penpri}.

\begin{figure}[htpb]\label{fig1-2-3dw}
\begin{center}
\includegraphics[width=6cm,height=5cm]{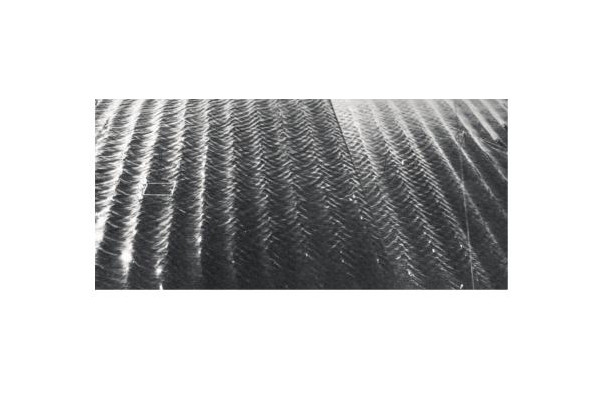}\\
\includegraphics[width=6cm,height=5cm]{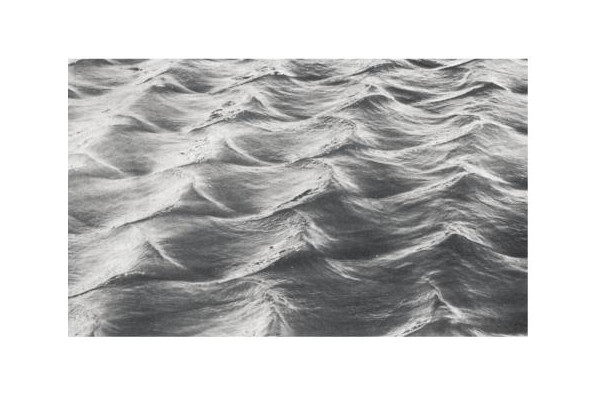}
\caption{(Top) Wave patterns in a basin showing two-dimensional subharmonic
instabilities ($ak=0.32$, $f=1.55$ Hz). (Bottom) A typical three-dimensional spilling
breakers in a basin. From Su {\em et al.} \cite{suetal82}.}
\end{center}
\end{figure}
  
Longuet-Higgins \cite{LH6,LH7} investigated, by combination of numerical analytical techniques, the stability of finite amplitude water waves to superharmonic and subharmonic two-dimensional disturbances.  His work  was an extension of results obtained by Zakharov and Benjamin \& Feir to finite amplitude waves for
disturbances of shorter wavelength.  Longuet-Higgins analysis \cite{LH6,LH7} confirmed Lighthill's prediction that the long wave instability is no longer present 
when the waves become very steep (that is when the steepness $ak\gtrsim 0.3$) and gave values for growth rates that agree well with the observations of Benjamin \& Feir, as well as Benjamin \cite{benj67} and Lake
\& Yuen \cite{lakeyuen}.  Longuet-Higgins also discovered that when the wave
is sufficiently steep, two-dimensional subharmonic disturbances of
twice the wavelength of the undisturbed wave become unstable and
have growth rates substantially larger than the type studied by
Lighthill, Zakharov and Benjamin \& Feir. Fig.~2 (left) shows such wave patterns up to 23 wavelengths. This photograph shows the first few waves contain small subharmonic disturbances which grow in size and height. These perturbations are three-dimensional, with wavelength of $\tf{1}{3}$ of original basic wave. Fig.~3 (right), on the other hand, depicts the three-dimensional spilling breakers due
to subharmonic instabilities in which the nature of the crescent-shaped breakers can clearly be seen. As Su {\em et al.} \cite{suetal82} commented, the three-dimensionality of the spilling breakers is an intrinsic characteristic of the three-dimensional subharmonic instabilities. 

\section{Three-dimensional water waves}

Following the advances mentioned above for two-dimensional water waves,
attention has been focused to three-dimensional propagating water waves, 
where an even greater richness of phenomena has been encountered \cite{SaffYuen80}.
We remark that  an important
distinction in three-dimensional water waves has to be made, namely between
{\em forced} and {\em spontaneous} three-dimensional waves.  The forced three-dimensional water waves is where the dependence upon the
second horizontal dimension is forced by boundary  or initial
conditions.  This is the case when one is concerned with studying
the effect of nonlinearity on the interaction of two equal but
non-parallel wave trains or the reflection of an obliquely
incident wave train on a wall \cite{rossaj}, see Fig.~3 (bottom).  As it has already been remarked, a special case when the waves are at an angle of $\tf{1}{2}\pi$ to each other  represents a two-dimensional standing wave \cite{rossaj}.

Forced three-dimensional waves, from a mathematical stand point, are essentially 
{\em superharmonic} modifications of the fundamental waves.  The basic linear state is given by \cite{sajrev}
\be
\eta(x,y,t)=a\cos[kx\cos\theta+ky\sin\theta-\omega t]+
a\cos[kx\cos\theta-ky\sin\theta-\omega t]\label{1.1}
\en
with $\omega=\Omega(k)$ being the linear dispersion relation for
waves with wavenumber $k$ where two wave trains make an angle
$2\theta$ with each other.  The case $\theta=0$ is the limit of
two-dimensional propagating waves, and the case
$\theta=\tf{1}{2}\pi$ is a two-dimensional standing wave.  The steady propagating
finite amplitude forced three-dimensional
waves of permanent form is a solution which may be expressed in the form
\be 
\eta(x,y,t)=\Ssf{m=0}{\infty}\Ssf{n=0}{\infty}a_{mn}\cos[mk\cos\theta(x-ct)]\cos[nk\sin\theta y]\label{1.2}
\en
where $c$ is the wave complex speed.
These wave profiles are commonly referred to as short crested waves and
were first studied by Fuchs \cite{fuchs} and Chappelear \cite{chapp}.  Perhaps the most important property of these waves is that they exist in the
infinitesimal limit, and consequently can be calculated formally by
expansions in powers of wave height $h$.  However, there are
serious concerns as to whether such expansions converge and thus the existence of
these steady short crested waves is still not fully determined. For standing waves,  
that is when $\cos\theta\rightarrow 0$ and with $c$ being finite, the first
comprehensive study was made by Penney \& Price \cite{penpri} and later on by Schwartz \&
Whitney \cite{schwhit}.  The existence of these waves are also uncertain. 

\begin{figure}[htpb]\label{fig3dw-a-b}
\begin{center}
\includegraphics[width=9cm]{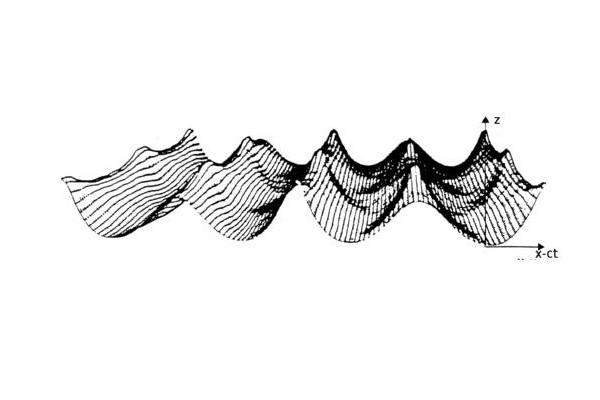}\\
\includegraphics[width=9cm]{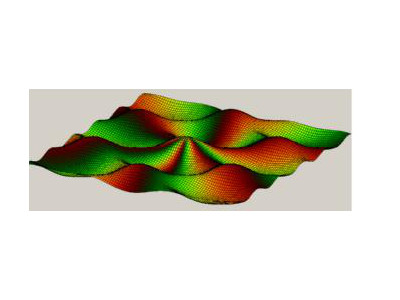}
\caption{(Top) A perspective plot of a spontaneous three-dimensional wave for $p=\tf{1}{2}$
and $q=1.2$. (Bottom) A perspective plot of a forced three-dimensional wave for $ak=0.1$.}
\end{center}
\end{figure}

In contrast, the spontaneous three-dimensional waves are completely different to that of forced three-dimensional waves because
these waves originate by instabilities or bifurcation of a uniform
two-dimensional wave train and, in general, they cannot have an
arbitrary small amplitude.  Mathematically speaking, they can be
described or interpreted as {\em subharmonic} bifurcations \cite{LH7},
where a two-dimensional wave of wavelength $2\pi/k$
\be
\os{\eta}(x,t)=\Ssf{\ell=0}{\infty}a_\ell\cos\ell k(x-ct)\label{1.3}
\en
bifurcates at a critical height into a steadily propagating
three-dimensional wave of the form
\be
\eta(x,y,t)=\os{\eta}(x,t)+\Ssf{\ell=0}{\infty}\Ssf{m=-\infty}{\infty}\Ssf{n=-\infty}{\infty}
A_{\ell,m,n}\cos[(\ell+mp)k(x-ct)+knqy]\label{1.4}
\en
where $p$ and $q$ are arbitrary real numbers with $0<p<1$.  However, if $p$
is an integer, these represent short crested waves.  The critical wave
height at which bifurcation occurs depends upon the values of $p$
and $q$.  The surface elevation, $\eta(x,y,t)$, given by (\ref{1.4}), 
is periodic in the transverse direction and having wavelength $2\pi/kq$.  
The longitudinal variation of these waves
can be thought of having wavelength $2\pi/kp$. Note incidentally, these waves are not exactly periodic unless $p$ is rational.  In the particular case when $p=1/2$ these waves
correspond to those whose wavelength is doubled in the direction of
propagation.

The existence of such waves was first demonstrated by Saffman \&
Yuen \cite{SaffYuPRL} for general nonlinear dispersive systems where the
nonlinearity is described by four wave interactions and their detailed 
calculations for water waves can be found in the paper Saffman \& Yuen \cite{SaffYuen80} 
using the Zakharov equation.  Saffman \& Yuen have pointed out that 
if the medium is isotropic, the bifurcation is degenerate and the solutions 
on the new branches can be either skew or symmetric. For skew branches
$
A_{\ell,m,n}\neq A_{\ell,m,-n}
$
and in this case the wave surface is not symmetric about the direction of
propagation.  On the other hand, for symmetric branches $A_{\ell,m,n}=A_{\ell,m,-n}$ and 
the surface is symmetric about the direction of propagation as in  short
crested waves. In this case, the surface elevation may be described by
\be
\eta(x,y,t)=
\os{\eta}(x,t)+
\Ssf{\ell=0}{\infty}\Ssf{m=-\infty}{\infty}\Ssf{n=0}{\infty}
A_{\ell,m,n}(p,q)\cos[(\ell+mp)(x-ct)]\cos(nqy)\label{1.6}
\en
with the corresponding velocity potential, satisfying equations (\ref{2.1}) below, 
is given by
\be 
\phi(x,y,z,t)=\os{\phi}(x,t)+\Ssf{\ell=0}{\infty}\Ssf{m=-\infty}{\infty}\Ssf{n=0}{\infty}
B_{\ell,m,n}(p,q)\exp(\omega_{\ell,m,n}y)\sin[(\ell+mp)(x-ct)]\cos(nqy)\label{1.6a}
\en
where $\omega_{\ell,m,n}=[(\ell+mp)^2+n^2q^2]$, and
both $A_{\ell,m,n}$ and $B_{\ell,m,n}$ are Fourier coefficients, see Section 3.
Solutions of this type have been calculated by Meiron {\em et al.} \cite{MeieEtAl} using the exact water wave equations (\ref{2.1}). A typical example is shown in Fig.~3 (top).

\begin{figure}[htpb]\label{fig5-3dw}
\begin{center}
\includegraphics[width=9cm]{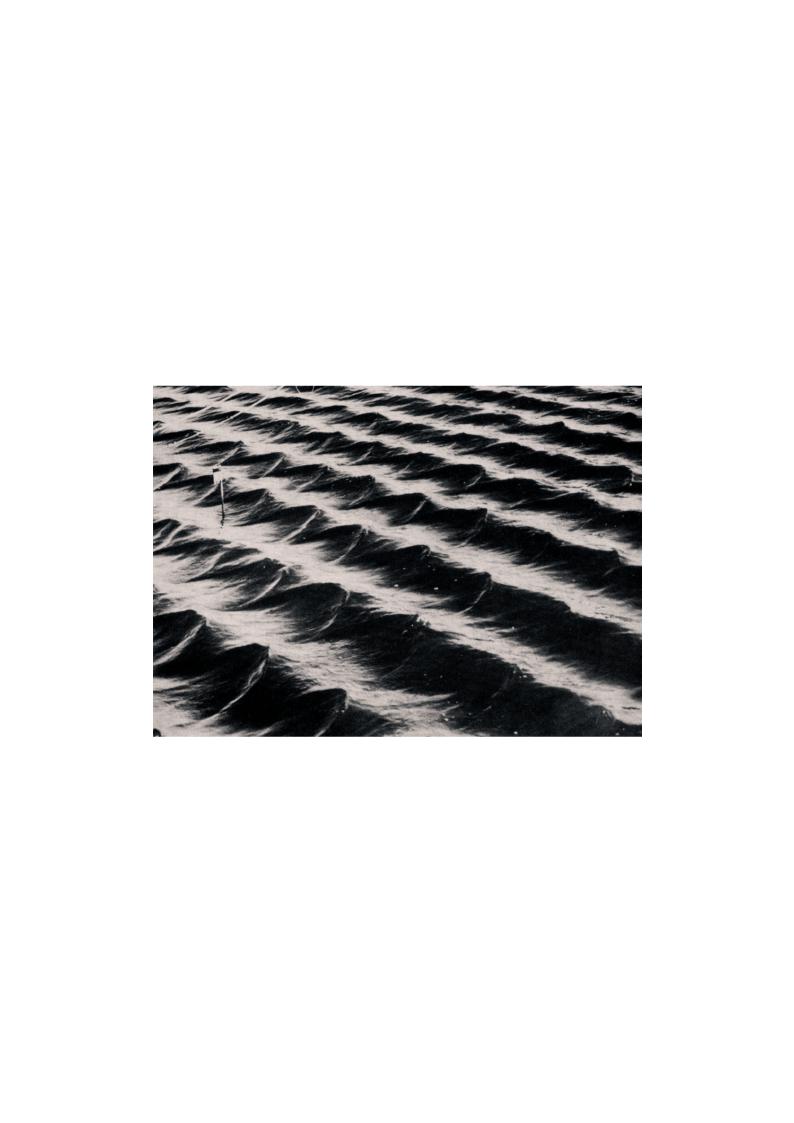}
\caption{Three-dimensional symmetric waves ($ak=0.33$, $f=1.2$ Hz) in a wide basin. 
From Su \cite{Su81}.}
\end{center}
\end{figure}

A typical experimental example of symmetric waves in a wide basin generated by a wavemaker with frequency $f=1.2$ Hz, wave steepness $ak=0.33$ and
wavelength $\lambda=1.08$ m is shown in Fig.~5.
We remark that the anti-symmetrical modes of skew waves
will lead to a branch of bifurcated solutions in which the surface
can be expressed as
\be
\eta(x,y,t)=\sum^{\infty}_{m=0}\sum^{\infty}_{n=-\infty}A_{m,n}\cos[\tf{1}{2}m(x-ct)+nqy]\label{4.1}
\en
with $A_{m,n} \neq A_{m,-n}$. These
skew waves have the property that a frame of reference can be chosen in
which the surface is stationary.   

\begin{figure}[htpb]\label{fig4-3dw}
\begin{center}
\includegraphics[width=10cm]{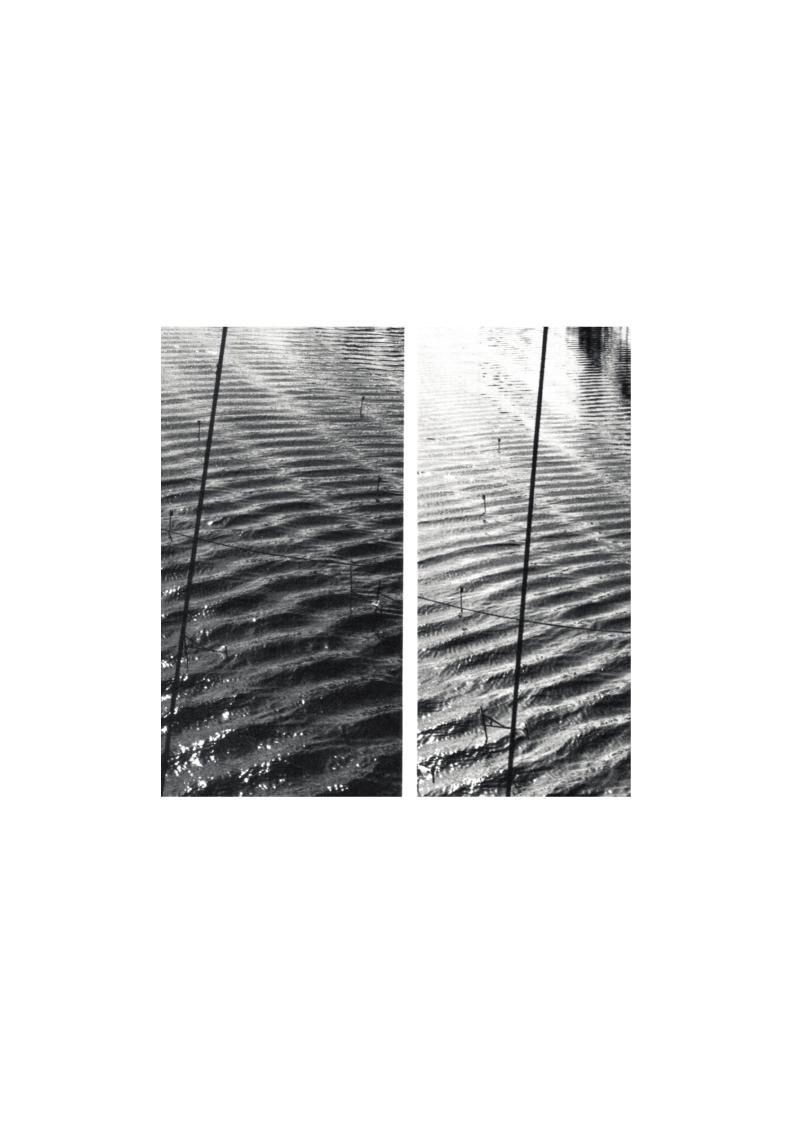}
\caption{Skew wave patterns for wave steepness $ak=0.17$. From Su \cite{suetal82}.}
\end{center}
\end{figure}

Fig.~6, on the other hand, depicts photographs of four skew patterns that propagate to the right side of the $x$-axis. These are three-dimensional wave-forms of the skew wave groups in which each pattern propagates at an angle $\alpha$ with respect to the
$x$-direction of the initial Stokes wave train. It is to be noted that skew waves are generally observed in a narrow range of steepness of about $0.16\leq ak\leq 0.18$. However, the predominantly two-dimensional envelope modulations of Benjamin-Feir instability are uniform over a wider range of steepness $0.1\leq ak\leq 0.25$ \cite{McLetal80, Su81}. This observation implies that the bifurcation rate for the skew waves is much lower than the growth rate of Benjamin-Feir instability for $ak<0.16$ and $ak>0.18$ \cite{Su81}.

The approach to the instability and shapes of
permanent waves, described above, relies on solving exact formulations, given by equations (\ref{2.1}).  An alternative approach, valid for weak nonlinearity and slow
modulations, is solving the cubic nonlinear Schr\"odinger equation exactly.

The aim of this paper is to classify various possible exact solutions of the elliptic ordinary differential equation that arises from nonlinear Schr\"odinger equation under the
decomposition given by (\ref{eq1}). To the best of our knowledge as yet no such
classification has been reported in the open literature. We anticipate the classification presented in this paper will assist in adopting physical solutions to nonlinear Schr\"odinger equation, for three-dimensional waves, amongst various possible ones.  

\section{Formulation of the problem}

The governing equations for an irrotational, inviscid, incompressible surface
gravity waves on deep water are given by
\be 
\begin{array}{lcl}
\nabla^2\phi=0, & & -\infty<z<\eta(x,y,t)\\
\\
\left.\begin{array}{l}
\PDD{\phi}{t}+\tf{1}{2}|\B{\nabla}\phi|^2+g\eta=0,\\
\\
\PDD{\eta}{t}+(\B{\nabla}\eta\B{\cdot\nabla}\phi)-\PDD{\phi}{z}=0,
\end{array}\right\} & & z=\eta(x,y,t)\\
\\
\B{\nabla}\phi\rightarrow 0 & & z\rightarrow -\infty\\
\end{array}\label{2.1}
\en 
where $g$ is the acceleration due to gravity, $\phi(x,y,z,t)$ is the
velocity potential, and $\eta(x,y,t)$ describes the free surface elevation.

Following \cite{zakharov} the surface elevation $\eta(\B{x}, t)$, where $\B{x}=(x,y)$, of weakly nonlinear deep-water gravity waves may be expressed as
\be 
\eta(\B{x},t)=\df{1}{2\pi}\Lint{-\infty}{\infty}\df{|\B{k}|^{\Quarter}}{2^{\Half}g^{\Quarter}}\left[B(\B{k},t){\rm e}^{\i(\B{k\cdot x}-\omega t)}+
B^*(\B{k},t){\rm e}^{-\i(\B{k\cdot x}-\omega t)}\right]\,{\rm d}\B{k}\label{2.2}
\en 
where superscript * denotes complex conjugates and $\omega$ is the wave frequency which satisfies the dispersion relation
\be 
\omega(\B{k})=\sqrt{g|\B{k}|}\label{2.3}
\en 
In equation (\ref{2.2}) $B(\B{k},t)$ is the time evolution of spectral
components of a weakly nonlinear system for dominating four-wave interactions 
and its governing equation is given by
\be 
\i\PDD{B(\B{k},t)}{t}=\Intn\!\!\Lint{-\infty}{\infty}\!\!\Intn T(\B{k},\B{k}_1,\B{k}_2,\B{k}_3)\delta(\B{k}+\B{k}_1-\B{k}_2-\B{k}_3)\exp\{\i[\omega(\B{k})+\omega(\B{k}_1)-\omega(\B{k}_2)
-\omega(\B{k}_3)]t\}& &\no\\
\times B^*(\B{k}_1,t)B(\B{k}_2,t)B(\B{k}_3,t)\,{\rm d}^3k_1\,{\rm d}^3k_2\,{\rm d}^3k_3\label{2.4}
\en 
where $\omega(\B{k})$ is the linear frequency  and the real interaction coefficient $T(\B{k},\B{k}_1,\B{k}_2,\B{k}_3)$, given by originally by Zakharov \cite{zakharov} and later, with some minor corrections, by Crawford {\em et al.} \cite{CrawEtAl}, characterize the properties of the system. For clarity we have listed these coefficients in the Appendix.

Zakharov \cite{zakharov} in his pioneering paper showed that equations (\ref{2.2})-(\ref{2.4})
yield the surface elevation for deep-water gravity wave (note, in his original
paper he also took into account the surface tension) may be reduced to
\be 
\eta(\B{x},t)=\Real\left\{A(\B{x},t){\rm e}^{\i(k_0-\omega_0t)}\right\}\label{2.5}
\en 
where $A(\B{x},t)=a(\B{x},t){\rm e}^{\i\theta(\B{x},t)}$ is the complex envelope
of the slowly modulated carrier wave propagating in the $\B{x}$-direction. The real envelope is given by $a(\B{x},t)$, and $\B{\nabla}_x\theta$ and 
$\partial\theta/\partial t$ represent, respectively, the modulation wave vector
and frequency. Zakharov showed that $A(\B{x},t)$ satisfies the two-dimensional
nonlinear Schr\"odinger equation
\be 
\i\left(\PDD{A}{t}+\df{\omega_0}{2k_0}\PDD{A}{x}\right)-\df{\omega_0}{8k_0^2}
\PDDT{A}{x}+\df{\omega_0}{4k_0^2}\PDDT{A}{y}-\tf{1}{2}\omega_0k_0^2|A|^2A=0
\label{2.6} 
\en 
Note incidentally, the complex envelope function $A(\B{x},t)$ is related to the Fourier
components $B(\B{k},\tau)$ through the equation
\be 
A(\B{x},t)=\df{\eps}{\pi}\df{k_0}{2\omega_0}\Lint{-\infty}{\infty}B(\B{k},t)
\exp\{\i(\B{k}-\B{k}_0)\B{\cdot x}-\i[\omega(\B{k})-\omega_0]t\}\,{\rm d}\B{k}\no 
\en 

Zakharov \cite{zakharov} also showed  that for oblique plane modulations, where $\xi=x\cos\alpha+y\sin\alpha$, equation (\ref{2.6}) may be reduced to one-dimensional nonlinear schr\"odinger equation
\be 
i\left(\PDD{A}{t}+\df{\omega_0}{2k_0}\cos\alpha\PDD{A}{\xi}\right)-\df{\omega_0}{8k_0^2}(1-\sin^2\alpha)
\PDDT{A}{\xi}-\tf{1}{2}\omega_0k_0^2|A|^2A=0
\label{2.6a} 
\en  
As was originally shown by Saffman \& Yuen \cite{SaffYuPhysFl} for $\alpha<\sin^{-1}(1/3)=19.47^\circ$, equation (\ref{2.6a}) has the following soliton solution 
\be 
A(\xi,t)=a_0\sech\left\{\df{k_0^2a_0}{\sqrt{1-3\sin\alpha}}\left(\xi-\df{\omega_0}{2k_0}t\cos\alpha\right)\right\}\exp\left(-\tf{1}{4}i\omega_0k_0^2a_0^2t\right)\label{2.6ab}
\en 
whose profile is depicted in Fig.~6 for $\alpha=\tf{1}{12}\pi$. We remark that for $\alpha >19.47^\circ$, there
are no steady solutions to equation (\ref{2.6a}) that decay as $|\xi|\rightarrow\infty$. Note that Saffman \& Yuen \cite{SaffYuPhysFl} state that no steady solution exist for $\alpha>35.26^\circ$ and they plot their result, using equation (\ref{2.6ab}), for the angle $\alpha=30^\circ$. This is an error because for $\alpha>19.47^\circ$ the term $1-3\sin\alpha$ will be negative and thus $A(\xi,t)$ will no longer be a $\sech$
profile but instead it will yield a periodic solution. Note, in this case $\sech(i\chi)=\sec(\chi)$, where $\chi$ represents the argument of $\sech$ in equation (\ref{2.6ab}). Fig.~6 (bottom) shows the plot of $\sec$ solution for values of arguement slightly below and above the range $-\tf{1}{2}\pi<\chi<\tf{1}{2}\pi$. This is because the $\sec$ solution of the equation equation (\ref{2.6ab}) becomes unbounded for $\chi=\tf{1}{2}(2n+1)\pi$ since at these values $\sec(\chi)$ is infinite.  

\begin{figure}[htpb]\label{soliton}
\begin{center}
\includegraphics[width=12cm]{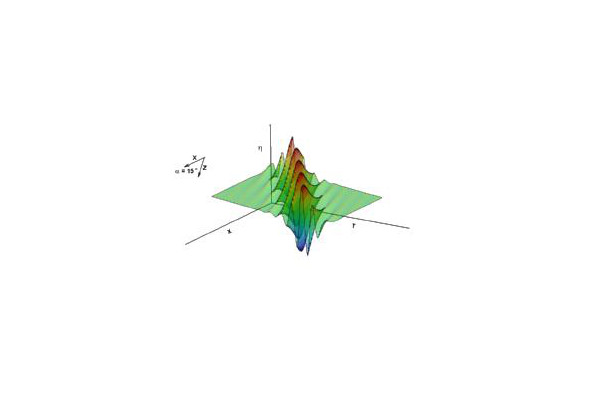}\\
\includegraphics[width=10cm]{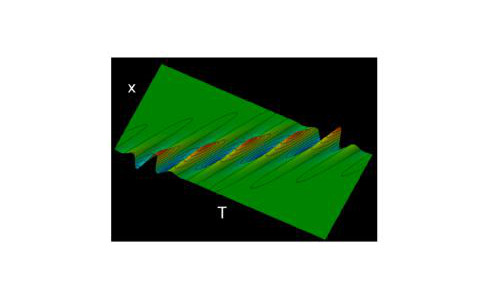}
\includegraphics[width=10cm]{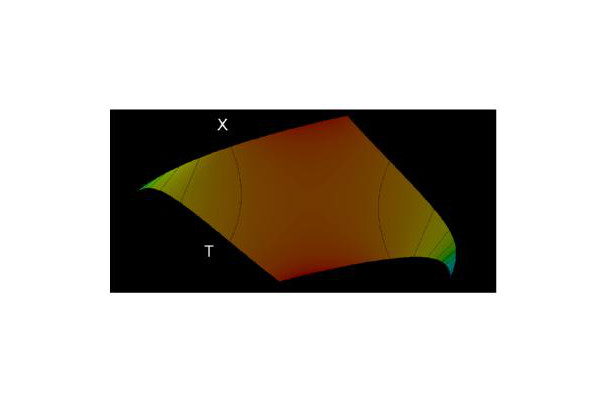}
\caption{(Top) Free surface elevation for a plane oblique solitary wave making an angle
of $\tf{1}{12}\pi$ with the carrier wave vector. (Middle) A different prospective of three
surface elevation as that of the top figure except for an angle $\tf{1}{15}\pi$. (Bottom)
is the same as the middle profile for an angle $\tf{1}{6}\pi$, where the $\sech$-profile becomes 
a $\sec$ profile. Note at values $\tf{1}{2}(2n+1)\pi$, where $n$ is an integer (including zero), 
the $\sec$-profile becomes infinite.}
\end{center}
\end{figure}

Following a common practice, we adopt the dimensionless variables
$$T=-\omega_0t,\quad X=2k_0\left(x-\df{\omega_0}{2k_0}t\right),\quad
Y=2k_0y, \quad \psi=k_0A/\sqrt{2}$$
and obtain the equation (\ref{2.6}) in non-dimensional form 
\be \label{nls}
i\left(\psi_T+\tf{1}{2}\psi_X\right)-\tf{1}{8}\psi_{XX}+\tf{1}{4}\psi_{YY}=\tf{1}{2}|\psi|^2\psi.\label{2.7}
\en

This equation has been used to study stabilities and
bifurcation of three-dimensional water waves. This two-dimensional form 
of nonlinear Schr\"odinger equation (\ref{2.7}) has attracted a great
deal of attention because it can be solved exactly by inverse
scattering and other techniques, but we emphasize that this equation has a rather physically
limited range of validity.  In the range of its
validity, however, it provides an easy way to produce all kinds of three-dimensional 
wave patterns.  
For example, substituting
\be \psi(X,Y,T)=f(Y)e^{\i(pX-cT)}\label{2.8}
\en
into equation (\ref{2.7}) we obtain
\be 
f_{YY} + \tf{1}{2}p^2f -2f^3 +2(2c-p)f = 0. \label{(2.9)} 
\en
Now, depending on the values of the parameters $p$ and $c$, and the
magnitude of $f$, many bounded solutions of this equation
exist (see Roberts \&
Peregrine \cite{RobPer} and Sections 4 and 5 below).  For example, there are stationary 
dislocation-type slip line solutions 
\be f = a \tanh(\sqrt{2}cY),\qquad p=-a^2,\label{2.10}
\en
which describes a surface in which the propagating wave has a
phase jump of $\pi$ across the line $Y = 0$ (see the case (2a)(ii) below). 
The problem with the nonlinear Schr\"odinger equation is that
it has an overabundance of solutions, and
it is not easy to decide which, if any, are of physical
significance.  There is also the mathematical problem of
determining if the nonlinear Schr\"odinger solutions are genuine in the sense
that there are limits, as the wavenumber of the modulation and the
amplitude go~to zero, of solutions of the exact equations, and not
the leading terms of expansions whose radius of convergence is
zero.

\section{Analytical solutions} 

To find analytical solutions to  \eqref{nls}  we employ the polar form of  ansatz 
\begin{equation}\label{eq1}
\psi(X,\z,T)=f(\z)e^{\i\theta}
\end{equation}
with  phase $\theta=kX-\omega T$,  $k$ is the wave number, $\omega$ is the angular frequency, and $f(\z)$ is the real magnitude of the field. 
Substituting the ansatz
into  \eqref{nls}, yields 
\begin{equation}\label{eq2}
f_{\z\z}+4 w(k)f=2f^3.
\end{equation}
By multiplying by $f_\z$ and integrating once we obtain the elliptic equation
\begin{equation}\label{eq3}
f_\z^2=f^4-4 w(k) f^2+\A
\end{equation}
where $\A$ is the integration constant that depends on the boundary conditions used, and
\begin{equation}\label{om}
w(k)=\omega-\frac k 2+\frac{k^2}{8}
\end{equation}
is the dispersion relation. We will make use of this disperion relation in order to classify the solutions and will be written 
as $\omega-\omega_0$ therein where $\omega_0=\tf{1}{2}k-\tf{1}{8}k^2$, see Fig.~7 which gives the graph of the  dispersion relation.

It is well known \cite{Wei, Whi} that
the solutions $ f(\z)$ of
\begin{equation}\label{eq4}
f_\z^2=p_4(f)
\end{equation}
where $p_4(f)$ is a quartic polynomial in $f(\z)$ can be expressed in terms of in  Weierstrass’ elliptic functions $\wp(\z;g_2,g_3)$ via
\begin{equation}\label{eq5}
f(\z)=f_0+\frac{ \sqrt{p_4(f_0)}\wp\rq{}(z)+\frac 1 2 {p_4}\rq{}(f_0)\Big(\wp(\z)-\frac {1}{24}{p_4}\rq{}\rq{}(f_0)\Big)+\frac{1}{24}p_4(f_0){p_4}^{(3)}(f_0)}{2\Big(\wp(\z)-\frac{1}{24}{p_4}\rq{}\rq{}(f_0)\Big)^2-\frac{1}{48}p_4(f_0)p_4^{(4)}(f_0)},
\end{equation}
where $f_0$ is not necessarily a root of $p_4(f)$, and $g_2,g_3$ are elliptic invariants of $\wp(Y)$.
Due to the biquadratic nature of $Q(f)$, an  analysis of the nature of solutions of  \eqref{eq5} is made, and we will show that solutions of \eqref{eq5} are reduced to solitary waves, periodic or Jacobi type of elliptic functions.

To see this we let $f^2(\z)=\zeta(\z)$ in \eqref{eq3}  which yields  the Weierstrass equation 
\begin{equation}\label{eq6}
\zeta_\z^2=4 \zeta^3-16 w(k) \zeta^2+ 4 \A\zeta
\end{equation}
Using a linear transformation $\zeta=\hat \zeta+\frac 43 w(k)$,  \eqref{eq6} can be written in normal form 
\begin{equation}\label{eq7}
\hat \zeta_\z^2=4\hat  \zeta^3-g_2 \hat \zeta -g _3=4(\hat \zeta-e_1)(\hat \zeta-e_2)(\hat \zeta-e_3)
\end{equation}

The germs of the Weierstrass equation \eqref{eq7}
are 
\begin{equation}\label{eq8}
\left.
\begin{aligned}
	g_2&=4 \left(\frac {16}{3}\omega^2(k)-\A\right)=2(e_1^2+e_2^2+e_3^2)\\
	g_3 &= \frac{16 \omega(k)}{3}\left(\frac {32}{9}\omega^2(k)-\A\right)=4(e_1e_2e_3)\\
\end{aligned}
\right\},
\end{equation}
and together with the modular discriminant 
\begin{equation}\label{del}
\Delta=g_2^3-27 g_3^2=16(e_1-e_2)^2(e_1-e_3)^2(e_2-e_3)^2
\end{equation}
are used to classify the solutions of \eqref{eq6}.
Also, $e_i$ are the three solutions to the cubic polynomial equation
\begin{equation}\label{eli}
p_3(s)=4s^3-g_2s-g_3=0,
\end{equation}
and are related to the two periods $\omega_{1,2}$ of the $\wp$ function for $e_{i}=\wp(\omega_{i}/2)$, and $\omega_3=-(\omega_1+\omega_2)$, see\cite{Steg}.

Since the solution of equation \eqref{eq7} is 
\begin{equation}\label{eq9}
\hat \zeta=\wp(\z;g_2,g_3)
\end{equation}
then the solution of equation \eqref{eq3} may be expressed as

\begin{equation}\label{eq10}
f(\z)=\pm \sqrt{\wp(\z;g_2,g_3)+\frac{4 \omega (k)}{3}}
\end{equation}

\section{Results}
We shall now proceed to classify the solutions of equation \eqref{eq3} case-by-case.
\begin{description}
\item [Case (1).] We consider the simpler case with zero boundary conditions, that is when $\A\equiv 0$. In this case equations \eqref{eq8} become
\be 
g_2=\frac{2^6}{3}\omega^2(k)\qquad\mbox{and}\qquad g_3 = \frac{2^9}{3^3}\omega^3(k)\label{eq11}
\en 
Also, in this case, \eqref{del} reduces to
 \begin{equation} \label{del2}
 \Delta  \equiv 0, 
\end{equation}  which implies that  $p_3(s)$ either has repeated root $e_i$ of multiplicity two ($m=2$) or three ($m=3$). 

\begin{description}
\item [Case (1a).] ($m=2$), we now let $e_1=e_2=u>0$ then $e_3=-2u<0$, hence
\be 
g_2=12u^2>0\qquad\mbox{and}\qquad g_3 =-8u^3<0\label{eq12}
\en 
Also, from equations \eqref{eq8} we see that 
 \begin{equation} \label{del2a}
u=-\frac 4 3 \omega(k)>0\quad \mbox{then}\quad \omega(k)<0\quad \mbox{which gives}\quad 
\omega <\omega_0=\frac k 2 -\frac{ k^2}{ 8},
\end{equation} 
see Fig.~7 (left), and hence, in this case \cite {Steg}, solution of \eqref{eq7} is given by 

\begin{equation} \label{eq13}
\wp_{1a}(\z;12u^2,-8u^3)=u+3u~ \mathrm{csch}^2(\sqrt{3u}\z).
\end{equation}

Using \eqref{del2}, \eqref{eq13} in \eqref{eq10}, we may express the solution of equation \eqref{eq3} as 
 \begin{equation} \label{eq14}
f_{1a}(\z)=2\sqrt{-\omega(k)} \left| \mathrm{csch}(2\sqrt{-\omega(k)}\z)\right|.
\end{equation}
For $k=1,~\omega=-1\rightarrow \omega(k)=-\frac {11}{8}$, equation \eqref{eq14} becomes 
 \begin{equation} \label{eq14a}
f_{1a}(\z)=\frac{\sqrt {22}}{2}\left|\mathrm{csch}\Big(\frac{\sqrt {22}}{2}\z\Big)\right|,
\end{equation}
 see Fig. \ref{FIG8} (blue).


\item [Case (1b).] ($m=2$), here we let $e_2=e_3=-u<0$ then $e_1=2u>0$, hence
\be 
g_2=12u^2>0\qquad\mbox{and}\qquad g_3 =8u^3>0\label{eq15}
\en 
Also, from equations \eqref{eq8} we have 
 \begin{equation} \label{del3}
u=\frac 4 3 \omega(k)>0\quad \mbox{then}\quad \omega(k)>0\quad \mbox{which gives}\quad 
\omega >\omega_0=\frac k 2 -\frac{ k^2}{8},
\end{equation} 
see Fig. 7 (center), thus in this case \cite {Steg} solution of equation \eqref{eq7} is given by 

\begin{figure}[htpb]\label{FIG7}
\begin{center}
\includegraphics[width=5.5cm]{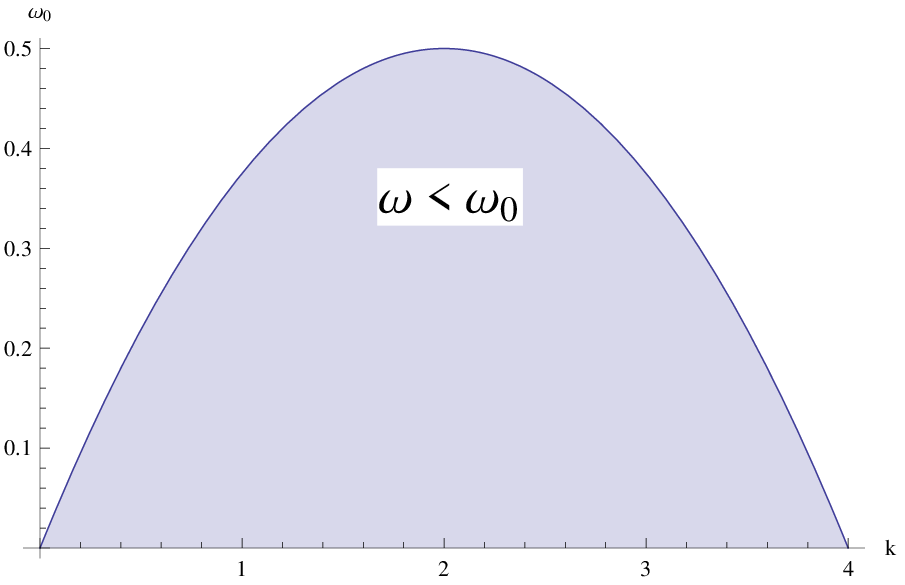}\includegraphics[width=5.5cm]{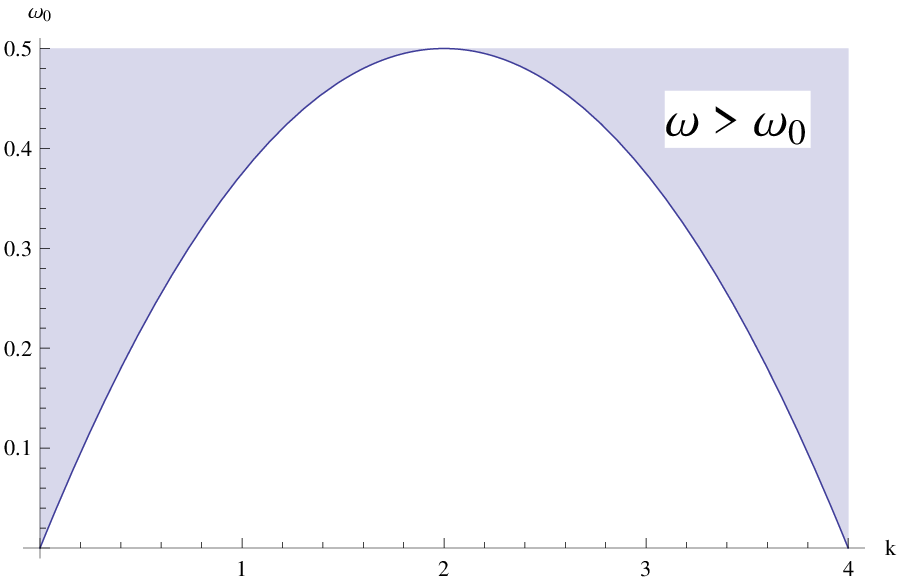}
\includegraphics[width=5.5cm]{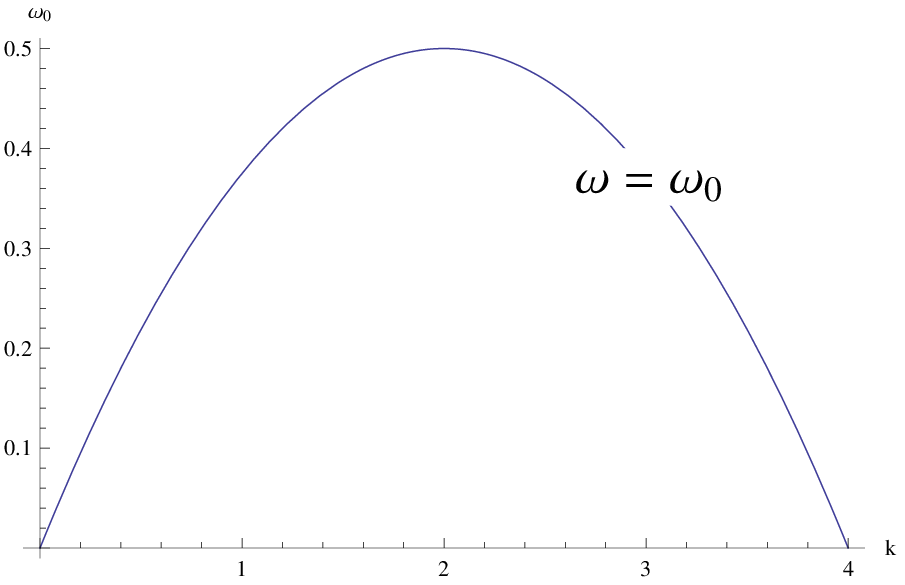}
\caption{Domain of solutions $f_{1a}$ (left), $f_{1b}$ (center), $f_{1c}$ (right), when  $\A=0$.}
\end{center}
\end{figure}

\begin{equation} \label{eq16}
\wp_{1b}(\z;12u^2,8u^3)=-u+3u~ \mathrm{csc}^2(\sqrt{3u}\z).
\end{equation}
Using \eqref{del3}, \eqref{eq16} in \eqref{eq10}, we may write the solution of \eqref{eq3} in the form
 \begin{equation} \label{eq17}
f_{1b}(\z)=2\sqrt{\omega(k)}\big| \mathrm{csc}(2\sqrt{\omega(k)}\z)\big|.
\end{equation}
For $k=1,~\omega=1$ then $\omega(k)=\frac 5 8$, equation \eqref{eq17} becomes 
 \begin{equation} \label{eq17a}
f_{1b}(\z)=\frac{\sqrt{10}}{2}\left|\mathrm{csc}\Big(\frac{\sqrt{10}}{2}\z\Big)\right|,
\end{equation}
 see Fig. \ref{FIG8} (red).

\item [Case (1c).] ($m=3$): here we obtain $e_1=e_2=e_3=0$, hence $g_2=g_3=0$ then $\omega(k)=0$ and this yields  $\omega=\omega_0=\frac k 2 -\frac{ k^2}{ 8}$. For $k=1,~\omega=\frac 3 8$ then $\omega(k)=0$. Hence
\begin{equation} \label{eq18}
\wp_{1c}(\z;0,0)=\frac{1}{\z^2},
\end{equation}
and 
\begin{equation} \label{eq19}
f_{1c}(\z)=\frac 1 \z
\end{equation}
see Fig. \ref{FIG8} (green).
\end{description}

\begin{figure}[htpb]
\begin{center}
\begin{tabular}{c}
\resizebox*{0.5\textwidth}{!}{\rotatebox{0}
{\includegraphics{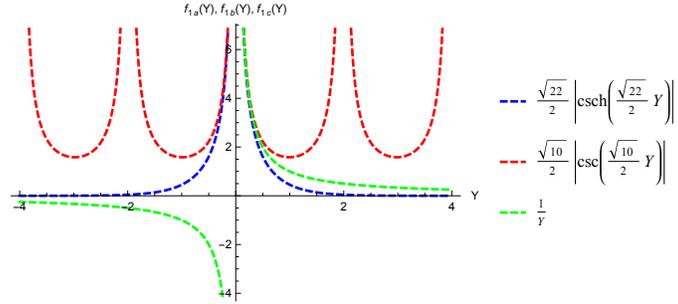}}}
\end{tabular}
\caption{Solutions when $\A=0$, for $k=1$, $\omega=-1$ (blue); $k=1$, $\omega=1$ (red); and  $k=1$, $\omega=\frac 1 3 $ (green).}\label{FIG8}
\end{center}
\end{figure}

\item [Case (2).] Now we consider the general case with nonzero boundary conditions, i.e., 
$\A\ne0$, then equation \eqref{eq3} is factored as

\begin{equation} \label{eq20}
f_\z^2=(f^2-a)(f^2-b),
\end{equation}
where
\begin{equation} \label{eq21}
a,b=2 \omega(k)\pm \sqrt{4 \omega^2(k)-\A}
\end{equation}
For real solutions of the amplitude $f$, we require $\A\le 4 \omega^2(k)$.
\begin{description}
\item [Case (2a).] For  $\A=4 \omega^2(k) \ne 0$ then $a=b=2 \omega(k)$, and equation \eqref{eq20} becomes

\begin{equation} \label{eq22}
f_\z=\pm (f^2-a)
\end{equation}

\begin{description}
\item [Case (2a)(i).] $\omega(k)<0$, then 
\begin{equation} \label{eq23}
f_{2ai}(\z)=\sqrt{-2 \omega(k)}\tan(\sqrt{-2 \omega(k)}\z).
\end{equation}
Now, for  $k=1,~\omega=-1$ then $\omega(k)=-\frac {11}{8}$ gives $\A=\frac {121}{16}$, and equation \eqref{eq23} becomes
\begin{equation} \label{eq23a}
f_{2ai}(\z)=\frac{\sqrt{11}}{2}\tan \Big( \frac{\sqrt{11}}{2}\z\Big),
\end{equation}
 see Fig. \ref{FIG9} (blue).
\item [Case (2a)(ii).] $\omega(k)>0$ then
\begin{equation} \label{eq24}
f_{2aii}(\z)=\sqrt{2 \omega(k)}\tanh(\sqrt{2 \omega(k)}\z).
\end{equation}
For  $k=1,~\omega=1\rightarrow \omega(k)=\frac 5 8 \rightarrow \A=\frac {25}{16}$ and equation \eqref{eq24} becomes
\begin{equation} \label{eq24a}
f_{2aii}(\z)=\frac{\sqrt{5}}{2}\tanh \Big( \frac{\sqrt{5}}{2}\z\Big),
\end{equation}
 see Fig. \ref{FIG9} (red).

\end{description}

\begin{figure}[htpb]
\begin{center}
\begin{tabular}{c}
\resizebox*{0.5\textwidth}{!}{\rotatebox{0}
{\includegraphics{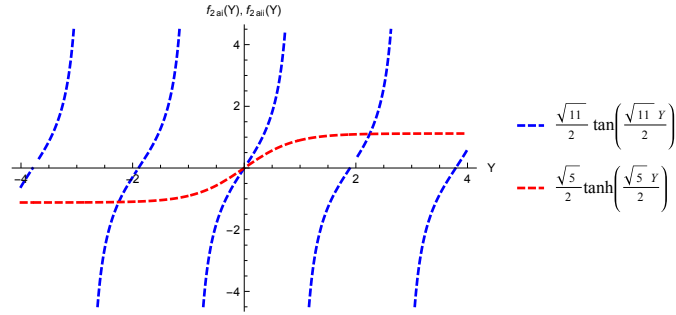}}}
\end{tabular}
\caption{Solutions when $\A=4\omega^2(k)$, for $k=1$, $\omega=-1$, $\A=\frac{121}{16}$  (blue);  and  $k=1$, $\omega=1$ (red),  $\A=\frac{25}{16}$.}\label{FIG9}
\end{center}
\end{figure}
\begin{figure}[htpb]\label{FIG7a}
\begin{center}
\includegraphics[width=12cm]{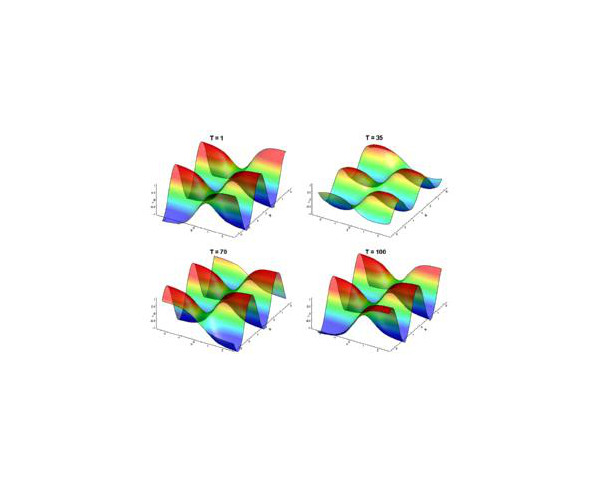}
\caption{Time series $\tanh$-solution of cubic nonlinear Schr\"odinger equation,                                                                                                                                                                                                                                                                                                                                                                                                                                                                                                                                                                                                                                                                                                                                                                                                                                                                                                                                                                                                                                                                                                                                                                                                                                                                                                                                                                                                                                                                                                                                                                                                                                                                                                                                                                                                                                                 for the same parameters as that in Fig 9.}
\end{center}
\end{figure}

\item [Case (2b).] Let  $\A<4 \omega^2(k)$, then there are two distinct roots $a,b$
which are given by
\be 
a=2 \omega(k)+\sqrt{4 \omega^2(k)-\A}\qquad\mbox{and}\qquad b=2 \omega(k)-\sqrt{4 \omega^2(k)-\A}
\label{eq25}
\en 
and we need to consider the following sub-cases.
\begin{description}
\item [Case (2b)(i).] $0<\A<4\omega^2(k)$, with $w(k)>0$ then $a=\hat{a}^2>0$, $b=\hat{b}^2>0$.
Hence, equation \eqref{eq3} becomes
\begin{equation} \label{eq26}
f_\z^2=(f^2-\hat{a}^2)(f^2-\hat{b}^2)
\end{equation}
In this case the solution of equation \eqref{eq26} is 
\be \label{eq26a}
f_{2bi}(\z)=\hat{b}~ \mathrm{sn} \Big(\hat{a} \z, \frac{\hat {b}^2}{\hat{a}^2}\Big)= \sqrt{2 \omega(k)-\sqrt{4 \omega^2(k)-\A}}\hspace*{3cm}& &\no\\
\times \mathrm{sn} \Big(\sqrt{2 \omega(k)+\sqrt{4 \omega^2(k)-\A}}\z, \frac{2 \omega(k)-\sqrt{4 \omega^2(k)-\A}}{2 \omega(k)+\sqrt{4 \omega^2(k)-\A}}\Big)& &
\en 
For $k=1,~\omega=1$ then $\omega(k)=\frac 5 8$. If we let $\A=1$ this gives $a=2, ~b=\frac 1 2$, and we thus have 
\begin{equation} \label{eq26aa}
f_{2bi}(\z)=\frac{\sqrt 2}{2}\mathrm{sn}\Big(\sqrt 2 \z,\frac 14\Big)
\end{equation}
 see Fig. \ref{FIG10} (blue).

\item [Case (2b)(ii).] $\A<0<4\omega^2(k)$, with $w(k)>0$ then $a=\hat{a}^2>0$, $b=-\hat{b}^2<0$. Hence, equation \eqref{eq3} becomes
\begin{equation} \label{eq27}
f_\z^2=(f^2-\hat{a}^2)(f^2+\hat{b}^2)
\end{equation}

Here  the solution of equation \eqref{eq27} is 
\be \label{eq27a}
f_{2bii}(\z)=\hat{a}~ \mathrm{nc} \Big(\sqrt{\hat{a}^2+\hat{b}^2}~\z, \frac{\hat {b}^2}{\hat{a}^2+\hat{b}^2}\Big)=\sqrt{2\omega(k)+\sqrt{4 \omega^2(k)-\A}}\hspace*{3cm}& &\no\\
\times \mathrm{nc} \Big(2\sqrt{\omega(k)}\z, \frac{2 \omega(k)-\sqrt{4 \omega^2(k)-\A}}{4 \omega(k)}\Big)& &
\en 

For $k=1,~\omega=1$ then $\omega(k)=\frac 5 8$. Letting $\A=-\frac{11}{16}$ gives $a=\frac{11}{4}, ~b=-\frac 1 4$, and we obtain
\begin{equation} \label{eq27aa}
f_{2bii}(\z)=\frac{\sqrt{11}}{2}\mathrm{nc}\Big(\frac{\sqrt{15}}{2} \z,\frac{1}{15}\Big)
\end{equation}
 see Fig. \ref{FIG10} (red).

\begin{figure}[htpb]
\begin{center}
\begin{tabular}{c}
\resizebox*{0.5\textwidth}{!}{\rotatebox{0}
{\includegraphics{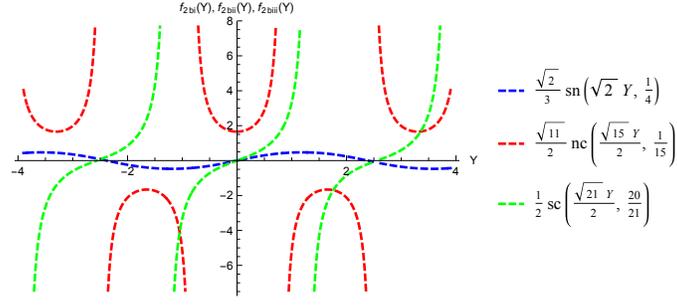}}}
\end{tabular}
\caption{Solutions when $\A<4\omega^2(k)$, for $k=1$, $\omega=1$, $\A=1$ (blue); $k=1$, $\omega=1$, $\A=-\frac{11}{16}$ (red); and $k=1$,  $\omega=-1$, $\A=\frac {21}{16}$ (green).}\label{FIG10}
\end{center}
\end{figure}

\begin{figure}[htpb]\label{FIG7b}
\begin{center}
\includegraphics[width=10cm]{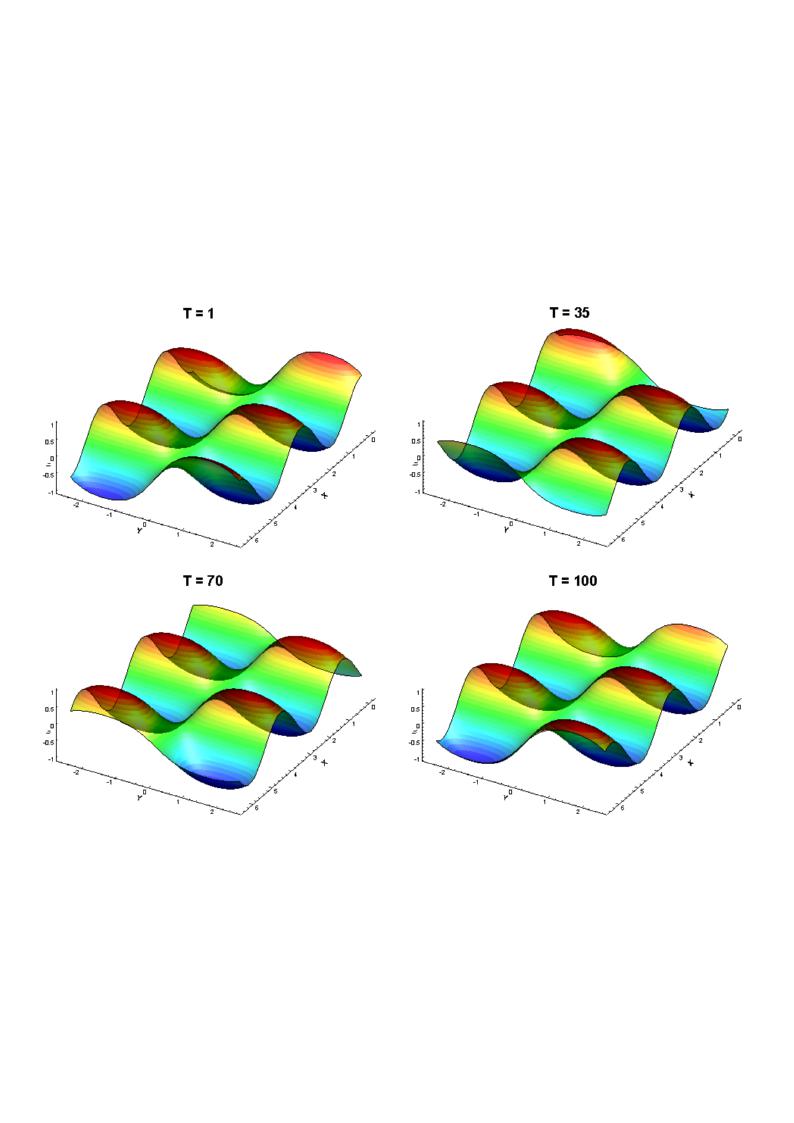}
\caption{Time series $\mathrm{sn}$-solution of cubic nonlinear Schr\"odinger equation,                                                                                                                                                                                                                                                                                                                                                                                                                                                                                                                                                                                                                                                                                                                                                                                                                                                                                                                                                                                                                                                                                                                                                                                                                                                                                                                                                                                                                                                                                                                                                                                                                                                                                                                                                                                                                                                 for the same parameters as that in Fig 11.}
\end{center}
\end{figure}

\item [Case (2b)(iii).] $0<\A<4\omega^2(k)$, with $w(k)<0$ then $a=-\hat{a}^2<0$, $b=-\hat{b}^2<0$.
Hence, equation \eqref{eq3} becomes
\begin{equation} \label{eq28}
f_\z^2=(f^2+\hat{a}^2)(f^2+\hat{b}^2)
\end{equation}

The solution of equation \eqref{eq28} is now given by 
\be \label{eq28a}
f_{2biii}(\z)=\hat{a}~ \mathrm{sc} \Big(\hat{b} \z, \frac{\hat {b}^2-\hat {a}^2}{\hat{b}^2}\Big)=\sqrt{2 \omega(k)+\sqrt{4 \omega^2(k)-\A}}\hspace*{3cm}& &\no\\ 
\times\mathrm{sc} \Big(\sqrt{2 \omega(k)-\sqrt{4 \omega^2(k)-\A}}\z, \frac{2\sqrt{4 \omega^2(k)-\A}}{2 \omega(k)-\sqrt{4 \omega^2(k)-\A}}\Big)& &
\en

For $k=1,~\omega=-1$ then $\omega(k)=-\frac{11}{8}$. Now if $\A=\frac{21}{16}$ gives $a=-\frac 1 4, ~b=-\frac{21}{4}$, and we get
\begin{equation} \label{eq28aa}
f_{2biii}(\z)=\frac 1 2 \mathrm{sc}\Big(\frac{\sqrt{21}}{2} \z,\frac{20}{21}\Big)
\end{equation}
 see Fig. \ref{FIG10} (green).

\end{description}
\end{description}
\end{description}

\section{Wave groups generated by wind}

Yuen \& Lake \cite{YuLake} proposed that a nonlinear wind-driven wave field may be characterized, to a first approximation, by a single nonlinear wave train. They based this proposal on experimental data obtained earlier for nonlinear deep-water wave trains without wind forcing. Using their laboratory experimental results they argued that the wind-wave interaction transports energy predominantly at a single speed corresponding to the group velocity, which is based on the dominant frequency corrected by wind-induced drift.

Based on their findings, they developed a model for the evolution of a nonlinear wave train in the absence of wind (which they argued is a satisfactory model) through the nonlinear Schr\"odinger equation
\be 
i\left(\PDD{\psi}{t}+\df{\omega_0}{2k_0}\PDD{\psi}{x}\right)-\df{\omega_0}{8k_0^2}\PDDT{\psi}{x}-\tf{1}{2}\omega_0\eps^2|\psi|^2\psi=0\label{G1}
\en 
where $\psi$ is the complex wave envelope, $\omega_0$ and $k_0$ are the carrier-wave frequency and wavenumber, and $\eps=k_0a_0$ is the inital steepness of the wave train. Similar to our analysis outlined in this paper, they related the free surface $\eta(x,t)$ to the wave envelope by the expression
$$\eta(x,t)=\Real\left\{\df{\eps}{k_0}\psi e^{\i(k_0x-\omega_0t)}\right\}.$$
Yuen \& Lake remarked that in a frame of reference moving with the group velocity $c_g=\omega/2k_0$ the variation of $\psi$ with $x$ and $t$ are of order $\eps k_0$ and $\eps^2\omega_0$, respectively, and is therefore slow compared to the oscillations of the dominant wave, characterized by $k_0$ and $\omega_0$.

Based on recent studies, e.g. \cite{SHD1, SHD2}, and our analysis here, it is very unlikely that Yuen \& Lake's model can adequately (if at all) represent nonlinear surface waves induced by wind, particularly in three dimensions.

A more consistent model, based on relatively a recent study by Leblanc \cite{labanc}, for three-dimensional deep water waves induced by wind forcing is to adopt the following two-dimensioanl version of nonlinear Schr\"odinger equation (cf. \cite{Sajhari})
\be 
2i\PDD{A}{\tau}+\mathscr{C}_1\PDDT{A}{\xi}+\mathscr{C}_2\PDDT{A}{\zeta}-\mathscr{C}_3|\psi|^2\psi=\omega(\alpha+i\beta)A\label{G2}
\en
where $\tau, \xi, \zeta$ are slow variables
$$\xi=\eps(x-c_gt),\qquad\zeta=\eps z,\qquad\tau=\eps^2t,$$
and $c_g=\partial\omega/\partial k$ is the group velocity. The coefficients $\mathscr{C}_i,\,\, i=1,2,3$ are given by
$$\mathscr{C}_1=\PDD{c_g}{k},\qquad\mathscr{C}_2=C_gk^{-1}\quad{\rm and}\quad\mathscr{C}_3=4k^4\omega^{-1}.$$
On the right-hand side of equation ({\ref{G2}), $\alpha$ and $\beta$ are the interfacial impedance, related to normal and tangential stresses at the air-sea interface \cite{Miles57}; $\beta$ is commonly known as the energy-transfer parameter.

\begin{figure}[htpb]\label{FIG7c}
\begin{center}
\includegraphics[width=8cm]{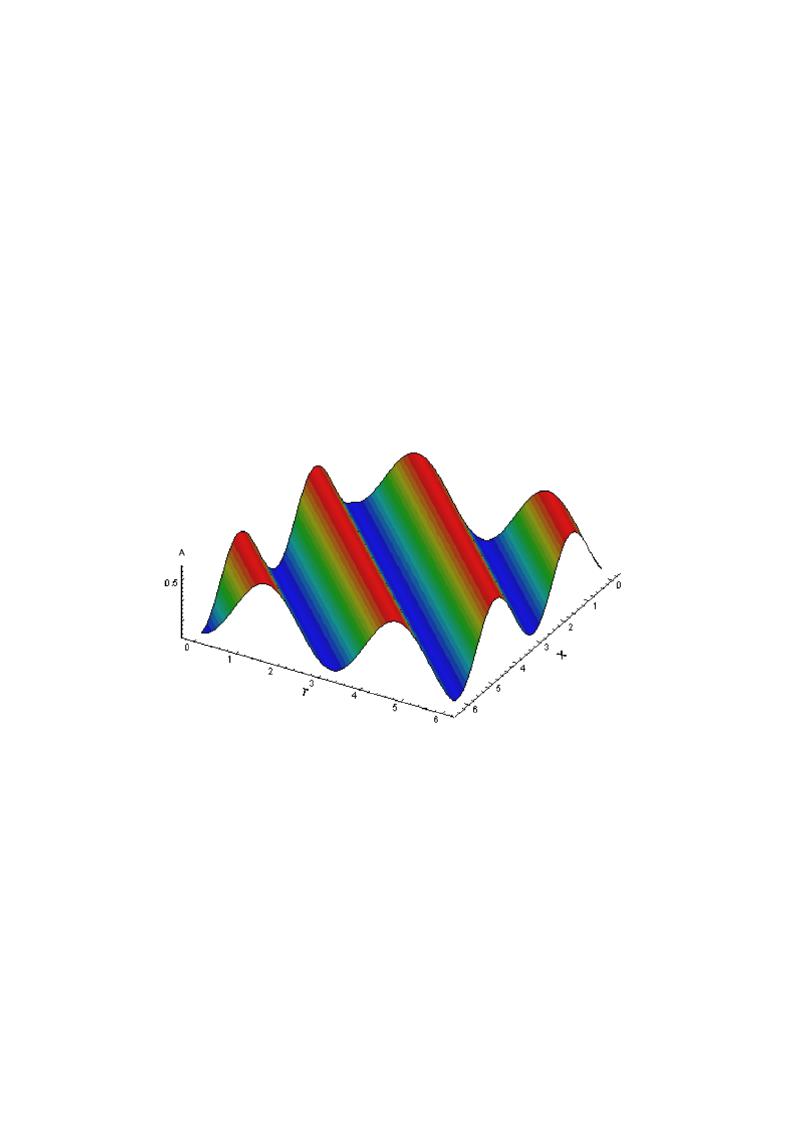}\quad\includegraphics[width=8cm]{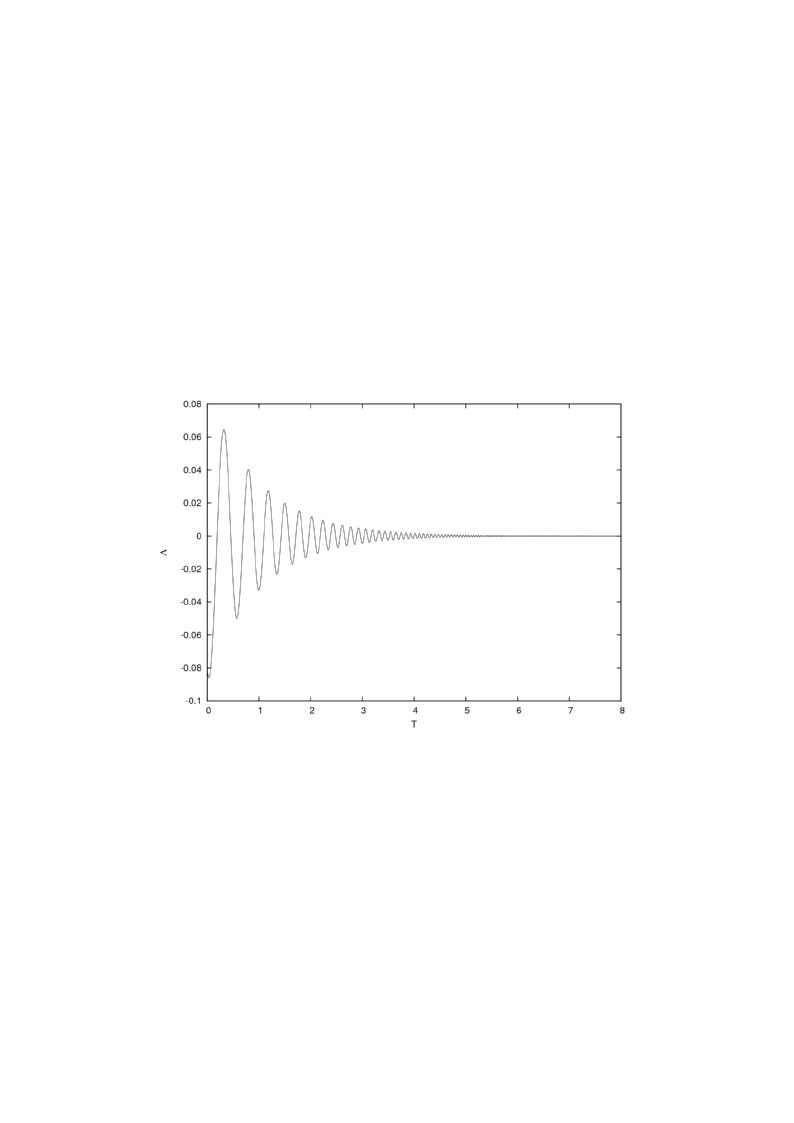}
\caption{(Left) Formation of a surface wave with wind forcing, calculated from $\beta$ by Miles' \cite{Miles57} formula, obtained from  $\mathrm{sn}$-solution of cubic nonlinear Schr\"odinger equation given in case (2b)(i)
above. (Right) Variation of $\Lambda$ with $T$ for $\beta<0$.}
\end{center}
\end{figure}

A solution of equation (\ref{G2}) may be sought in the form
$$A(\xi,\zeta,\tau)=(\omega/2k^2)\psi(X,Z,T)e^{-i\alpha T/2}$$
where $T=\omega\tau$, $X=k\xi$ and $Z=k\zeta$.
An exact homogeneous solution of (\ref{G2})  maybe obtained by writing $ \psi_s(T)=\mathscr{R}_se^{i\Theta_s}$ \cite{labanc} where
\be 
\mathscr{R}_s(T)=\mathscr{R}_0e^{\beta T/2},\qquad\Theta_s(T)=\Theta_0-\mathscr{R}_s^2/2\beta\label{G3}
\en 
where the suffix zero indicates the initial state. The solution corresponding to a spatially uniform plane Stokes wave is then obtained by letting $\beta\rightarrow 0$ and in this limit $\mathscr{R}_s\rightarrow\mathscr{R}_0e^{i(\Theta_0-\mathscr{R}_0^2T/2)}$.
On the other hand $\psi_s$ for an inhomogeneous solution is of the form $\psi_s(T)=e^{iKX+LZ}$ provided $K^2=2L^2$.


Following Leblanc \cite{labanc}, we seek a perturbation to the Stokes solution by writing
$$\psi=\mathscr{R}_s(1+\mu)e^{i(\Theta_s+\varphi)}$$ and expressing $\mu$ and $\varphi$ in the form
$$[\mu,\varphi]=[\Lambda(T),\Phi(T)]\cos(KX+LZ)$$
where $\Lambda$ satisfies the following ordinary differential equation
\be 
\Lambda''+\gamma(\gamma-\mathscr{R}_s^2)\Lambda=0,\quad\gamma=\tf{1}{8}(K^2-2L^2);\quad(\,\,\,)'\equiv {\rm d}/{\rm d}T\label{G4}
\en 

As in the analysis of Leblanc \cite{labanc} for two-dimensional waves, we see that in the absence of wind forcing ($\beta=0$) the solution grows exponentially provided $0<\gamma<\mathscr{R}_0^2$. This represents the three-dimensional Benjamin-Feir instability and is in direct contradiction with Yuen \& Lake \cite{YuLake}. However, when wind forcing is present ($\beta\neq 0$), it is easy to show the differential equation (\ref{G4}) admits two linearly independent solutions in terms of modified Bessel function $I_{\pm\nu}(f)$ of complex order $\nu=2i\gamma\beta^{-1}$ where $f(T)=2\gamma^{\Half}\mathscr{R}_s\beta^{-1}$. Thus we see that the decaying ($\beta<0$) solution of equation (\ref{G4}) oscillates in time and eventually damps due to viscous dissipation. On the other hand, the nondissipative solution ($\beta>0$) either decays if $\gamma<0$ or grows superexponentially \cite{labanc} when $\gamma>0$. In the latter case the asymptotic behaviour of equation (\ref{G4}) is given by \cite{labanc} 
$$\Lambda(T)\sim\df{e^{f(T)}}{\sqrt{2\pi f(T)}}\qquad\mbox{as $T\rightarrow\infty$.}$$
This results seems to suggest the three-dimensional waves for which $\beta>0$ suffer superharmonic instability initially but they are then suppressed by subharmonic instability. However, as yet no experimental evidence has been reported to support or reject this conjecture.

\begin{figure}[htpb]\label{middle}
\begin{center}
\includegraphics[width=10cm]{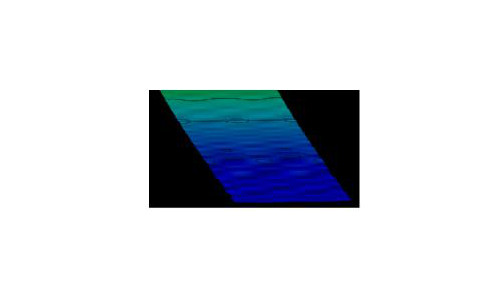}\\
\includegraphics[width=10cm,height=4.5cm]{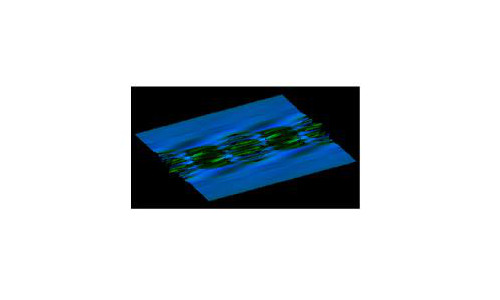}\\
\caption{(Top) Formation of wave groups which gradually grow in time.
(Bottom) Formation of wave groups with initial growth and subsequent decay in time.}
\end{center}
\end{figure}


\section{Concluding Remarks}

In this paper we have  reviewed several types of three-dimensional waves on deep-water. We have identified that three dimensionality of surface waves are essentially of three kinds namely those of oblique, forced and  spontaneous type. Although perturbation techiques can be used to describe these waves, we have shown an alternative formulation, through cubic nonlinear Schr\"odinger (NLS) equation, which in some cases may be superior. 
It is shown that, when adopting  the alternative approach the solution of NLS equation  must be carefully classified since this equation has overabundance of solutions and it is often difficult to decide which, if any, have physcial significance. Here, we have obtained various 
periodic solutions of the  NLS  equation using Weierstrass elliptic $\wp$ functions. It is shown the classification of solutions depends on the boundary conditions, wavenumber and frequency. We have demonstrated that in certain cases the solutions are of  solitary type, or simply periodic, while in other cases it is shown that solutions can be expressed in terms of Jacobi elliptic functions. For the formation of a group of waves,  analytical solution of forced (or inhomogeneous) NLS equation, which arises from  wind forcing, is found whose that clearly shows group of waves can form on the surface of deep water similar to those that are commonly observed in the ocean. In this case the depencency on the energy-transfer parameter, from wind to waves, is clearly a very influential factor, making either the groups of wave to grow initially and eventually dissipate or simply decay or grow in time.

\appendix
     
\section{Second and third order interaction coefficients}

Adopting the short-hand notation $T(k,k_1,k_2,k_3)=T_{0,1,2,3}$ we have
\be 
T_{0,1,2,3}=&-&\df{2\V^{(-)}_{3,3-1,1}\V^{(-)}_{0,2,0-2}}{\omega_{1-3}-\omega_3+\omega_1}-
\df{2\V^{(-)}_{2,0,2-0}\V^{(-)}_{1,1-3,3}}{\omega_{1-3}-\omega_1+\omega_3}-
\df{2\V^{(-)}_{2,2-1,1}\V^{(-)}_{0,3,0-3}}{\omega_{1-2}-\omega_2+\omega_1}\no\\
&-&\df{2\V^{(-)}_{3,0,3-0}\V^{(-)}_{1,1-2,2}}{\omega_{1-2}-\omega_1+\omega_2}-
\df{2\V^{(-)}_{0+1,0,1}\V^{(-)}_{2+3,2,3}}{\omega_{2+3}-\omega_2-\omega_3}-
\df{2\V^{(+)}_{-2-3,2,3}\V^{(+)}_{0,1,-0-1}}{\omega_{2+3}+\omega_2+\omega_3}\no\\
&+&\W_{0,1,2,3}\no 
\en 
Here, the second order interaction coefficients $\V^{(\pm)}_{0,1,2}$ are given by
\be 
\V^{(\pm)}_{0,1,2}=\df{1}{8\sqrt{2}\pi}\left\{(\B{k}_0\B{\cdot k}_1\pm k_0k_1)
\left(\df{\omega_0\omega_1}{\omega_2}\df{k_2}{k_0k_1}\right)^{\Half}+
(\B{k}_0\B{\cdot k}_2\pm k_0k_2)
\left(\df{\omega_0\omega_2}{\omega_1}\df{k_1}{k_0k_2}\right)^{\Half}+
(\B{k}_1\B{\cdot k}_2\pm k_1k_2)
\left(\df{\omega_1\omega_2}{\omega_0}\df{k_0}{k_1k_2}\right)^{\Half}\right\}\no 
\en 
where $k_i=|\B{k}_i|$ and $\omega_i=\omega(k_i)$.

Similarly, writing the short-hand notation $\W_{0,1,2,3}=\W(\B{k},\B{k}_1,\B{k}_2,\B{k}_3)$
we have for the third order interaction coefficients
\be 
\W_{0,1,2,3}=\U_{-0,-1,2,3}+\U_{2,3,-0,-1}-\U_{2,-1,-0,3}-
\U_{-0,2,-1,3}-\U_{-0,3,2,-1}-\U_{3,-1,2,-0}\no 
\en 
where
\be 
\U_{0,1,2,3}=\df{1}{64\pi^2}\left(\df{\omega_0\omega_1}{\omega_2\omega_3}k_0k_1k_2k_3
\right)^{\Half}\left\{2(k_0+k_1)-k_{1+3}-k_{1+2}-k_{0+3}-k_{0+2}\right\}\no 
\en 
with $k_{i\pm j}=|\B{k}_i\pm\B{k}_j|$ and $\omega_{i\pm j}=\omega(k_{i\pm j})$.

\end{document}